\documentclass[12pt,english]{article}

\usepackage[english]{babel}
\usepackage[utf8]{inputenc}
\usepackage{natbib,authblk,bbm,bm}
\usepackage{amsmath,setspace,geometry,lineno, amsfonts}
\usepackage{graphicx}
\usepackage[colorinlistoftodos]{todonotes}
\usepackage{fullpage}
\usepackage{tikz}
\usetikzlibrary{fit}
\usetikzlibrary{calc}
\usetikzlibrary{decorations.pathreplacing}
\usetikzlibrary{arrows.meta}
\tikzset{>={Latex[width=1.5mm,length=1.5mm]}}
\title{Multi-scale modeling of animal movement and general behavior data using hidden Markov models with hierarchical structures} 
\date{}

\author{Vianey Leos-Barajas$^{1*}$, Eric J. Gangloff$^1$, Timo Adam$^2$, Roland Langrock$^2$, Floris M.\ van Beest$^3$, Jacob Nabe-Nielsen$^3$ and Juan M. Morales$^4$\\ \small{$^1$Iowa State University, USA\\ $^2$Bielefeld University, Germany\\ $^3$Aarhus University, Denmark\\ $^4$INIBIOMA-CONICET, Argentina}}

\begin{document}
\begin{spacing}{1.9}

\maketitle

\begin{abstract}
Hidden Markov models (HMMs) are commonly used to model animal movement data and infer aspects of animal behavior. An HMM assumes that each data point from a time series of observations stems from one of $N$ possible states. The states are loosely connected to behavioral modes that manifest themselves at the temporal resolution at which observations are made. However, due to advances in tag technology, data can be collected at increasingly fine temporal resolutions. Yet, inferences at time scales cruder than those at which data are collected, and which correspond to larger-scale behavioral processes, are not yet answered via HMMs. We include additional hierarchical structures to the basic HMM framework in order to incorporate multiple Markov chains at various time scales. The hierarchically structured HMMs allow for behavioral inferences at multiple time scales and can also serve as a means to avoid coarsening data. Our proposed framework is one of the first that models animal behavior simultaneously at multiple time scales, opening new possibilities in the area of animal movement modeling. We illustrate the application of hierarchically structured HMMs in two real-data examples: (i) vertical movements of harbor porpoises observed in the field, and (ii) garter snake movement data collected as part of an experimental design.  
\end{abstract}


\section{INTRODUCTION}

Hidden Markov models (HMMs) and related state-switching models are prevalent in the field of animal movement modeling, where they provide a flexible framework to infer aspects of animal behavior from various types of movement data \citep{mor04, pat09, lan12, lan14}. They are very natural models for time series data related to animal movement, as they account for the serial dependence typically observed, yet also allow to (loosely) connect each observation to distinct underlying behavioral modes. A basic HMM for movement data consists of two stochastic processes: an observed movement process and an underlying state process, the latter of which can be related to distinct behavioral modes, at least in the sense of serving as a proxy of the actual behavioral process \citep{pat09,lan12}. Applications of HMMs to movement data often focus on investigating the effect of individual and environmental covariates on state occupancy, and thus ultimately on the dynamics of the variation in behavioral modes in response to internal and external drivers. 

Generally, movement data are analyzed such that the observation process is assumed to stem from a single (behavioral) state process. It may however be the case that there are two (connected) behavioral processes that occur at distinct time scales. For instance, so-called hierarchical HMMs have been used to process data on handwriting in order to distinguish between distinct letters but also to recognize a word, defined as a sequence of written letters \citep{fin98}. However, these versatile extensions of HMMs have not yet been applied to movement data, even in light of the intuitive idea that distinct behaviors manifest themselves at different time scales (hereafter referred to as multi-scale behaviors). A motivating example to have in mind is a central-place forager such as the southern elephant seal. These animals exhibit large-scale migration movements (from land colonies to either the sea ice zone around Antarctica or into open-ocean pelagic zones, and back), but also movement patterns where much more frequent changes take place between behavioral modes, e.g.\ ``foraging'' and ``resting'' modes \citep{mic16}. The modeling framework we propose regards such data as stemming from two behavioral processes, which operate on different time scales: the first process determines the behavioral mode at the cruder time scale (e.g., whether or not an elephant seal is performing a migratory trip, and also what kind of migratory trip), while the second process, at the finer time scale, determines the behavioral mode {\it nested within} the large-scale mode (e.g., whether an elephant seal is resting or foraging, given it is close to the sea ice zone, or whether it is traveling or foraging, given it is on a migratory trip). 

For multi-scale modeling of animal movement data, we propose an extension to the standard HMM that allows for a hierarchical state process, where two (or more) different Markov chains, operating at different time scales, will be tied together. To illustrate the application of hierarchically structured HMMs in a real-data setting, we model vertical movements of a harbor porpoise ({\it Phocoena phocoena}) throughout its natural habitat in the northeastern part of the North Sea. While the data were collected at a dive-by-dive resolution, the aim here is to infer dive patterns at two different temporal scales: an hourly scale to infer the general behavioral mode (e.g.\ resting or traveling), which may persist for a large number of consecutive dives, and a fine-scale process to infer more nuanced state transitions at a dive-by-dive resolution given the general behavioral mode. As a second real-data example, we model baby garter snake ({\it Thamnophis elegans}) movement data produced in a controlled experimental design context. The hierarchically structured HMM has two Markov chains, where one Markov chain models the transitions among three types of movements (distance traveled in 1/2 s) and the second Markov chain models transitions across the six time series produced per snake. As this group of garter snakes was assigned to the control group, indicating no treatment effect, we use the second Markov chain to investigate personality and repeatability in their movement patterns. That is, we attempt to answer if the garter snakes adapt their general movement strategies or if they have tendencies to exhibit the same general movement pattern across multiple time series.
 
A conceptual challenge with HMMs, and in fact any discrete-time models for behavioral data, is that the temporal resolution of the observations being analyzed (e.g., hourly, daily, etc.) determines what kind of behaviors may be inferred at all. Strictly speaking, this is not a problem arising from the model applied, but rather from the sampling protocol, i.e., the data. For instance, \cite{tow16} processed white shark location data, collected every five minutes, into distance traveled and turning angle and subsequently connected each bivariate observation to ``area-restricted search'' and ``transiting'' behavior. If the shark's location were observed once per day, we could not infer the same behaviors as switches between these behavioral modes occur at a much finer temporal scale. The hierarchically structured HMMs will not solve the conceptual challenges associated with data processing or data collection required to infer multi-scale behaviors. However, it does offer new opportunities in the analysis of animal movement data, allowing for identification of general behavioral patterns that are a composition of fine-scale observations and inferences to be made at multiple time-scales.

\section{HIDDEN MARKOV MODELS WITH HIERARCHICAL STRUCTURES}

In Section \ref{framework} we first detail the basic HMM framework in order to introduce the necessary notation that will be used throughout the paper. In Section \ref{hierarchical}, we introduce the hierarchical model formulation, distinguishing between two types of latent states, production states and internal states, which occur at distinct time scales.

\subsection{BASIC HMM FRAMEWORK}\label{framework}

A basic HMM is composed of two stochastic processes: an observable state-dependent process $\{Y_t\}_{t=1}^T$ and an unobservable state process $\{S_t\}_{t=1}^T$ taking on a finite number of states. Here we call the state a {\it production state} (as it produces an observation), in order to differentiate it from other forms of the latent states which we introduce in Section \ref{hierarchical}.
As is general practice, we assume a first-order Markov process at the production state level, such that $S_t$, the production state at time $t$, given the states at all previous times, will only depend on $S_{t-1}$. We further assume $Y_t$, $t=1,\ldots,T$, to be conditionally independent of past and future observations and production states, given the production state $S_t$, such that the production states effectively select from which of finitely many possible distributions each observation is drawn. Due to the Markov property, the evolution of the production states over time is governed by the transition probability matrix (t.p.m.), $\boldsymbol{\Gamma}=(\gamma_{ij})$, where $\gamma_{ij} = \text{Pr}(S_t=j|S_{t-1}=i)$ for $i,j = 1,...,N$, with $N$ denoting the number of production states. The initial distribution, $\boldsymbol{\delta}$, is a vector of probabilities with entries $\delta_i = \text{Pr}(S_1 = i)$, of the first observation $y_1$ belonging to one of the $N$ production states. It is common to assume the initial distribution to be the stationary distribution, defined as the solution to $\boldsymbol{\Gamma}\boldsymbol{\delta} = \boldsymbol{\Gamma}$. However, $\boldsymbol{\delta}$ can also be estimated. In order to ensure identifiability when estimating the entries of the t.p.m., we map the entries of each row onto the real line with the use of the multinomial logit link and set the diagonal entries of the matrix as the reference categories in the following manner,  
\begin{equation*}
\gamma_{ij} = \dfrac{\exp(\eta_{ij})}{\sum_{k=1}^N \exp(\eta_{ik})}, \qquad \text{where} \qquad
\eta_{ij} = 
\begin{cases}
	\beta^{(ij)}  & \text{ if } i \neq j;\\
	0 & \text{ otherwise}.
\end{cases}
\end{equation*}
We similarly use a multinomial logit link transformation for the initial distribution, if estimated rather than assumed to the be the stationary distribution. The state-dependent distributions for $Y_t$ will be represented in terms of probability density or mass functions $f(y_t | S_t=i) = f_i(y_t);  \, \, \, \, i=1, \ldots, N$. If the observations are multivariate, in which case we write $\mathbf{Y}_t=(Y_{1t},\ldots,Y_{Rt})$, we can either formulate a joint distribution $f_i(\mathbf{y}_t)$ or assume contemporaneous conditional independence by allowing the joint distribution to be represented as a product of marginal densities, $f_i(\mathbf{y}_t)$=$f_i^1(y_{1t})f_i^2(y_{2t})\cdots f_i^R(y_{Rt})$. While parametric families are usually chosen for the $f_i$, such as a Gaussian or gamma distribution, we can also estimate the distribution nonparametrically by expressing it as a linear combination of a large number of basis functions \citep{lan15}.
The likelihood of an individual time series can be expressed concisely as a matrix product,
\begin{equation}\label{basiclikelihood}
L_p(y_1,\ldots,y_T) = \boldsymbol{\delta}^{T} \mathbf{P}(y_1) \prod_{t=2}^T \boldsymbol{\Gamma} \mathbf{P}(y_t) \mathbf{1},
\end{equation}
where $\mathbf{P}(y_t) = \text{diag} \left( f_1 (y_{t}), \ldots, f_N (y_{t}) \right)$, $\boldsymbol{\delta}$ denotes the initial distribution, and $\mathbf{1}$ is a column vector of ones. Calculation of the matrix product given above — the computational cost of which notably is only linear in $T$ — is equivalent to applying the forward algorithm, which is an efficient recursive scheme for calculating the likelihood of an HMM \citep{zuc16}.

\subsection{EXTENSION TO ALLOW FOR HIERARCHICAL STRUCTURES}\label{hierarchical}

The framework for the basic HMM accounts for switches at the production state level. 
In a movement modeling analysis, the production states are generally thought to be proxies for behavior occurring at the time scale at which the data were collected (or processed). However, as outlined in the elephant seal example in the introduction, production states alone may not be sufficient to encompass complex multi-scale behavioral processes. More specifically, there may be crude-scale behavioral processes (e.g.\ migration) that manifest themselves as a sequence of production states and associated observations. Intuitively, we would then connect a behavior occurring at a cruder time scale to one of $K$ {\it internal states}, such that each internal state generates a distinct HMM, with the corresponding $N$ {\it production states} producing the actual observations.

Akin to the basic HMM framework, we can think of a fine-scale sequence of observations, $\mathbf{y}_{m} = (y_{1,m}, \ldots, y_{T,m}) $ — with one such sequence for each $m=1,\ldots, M$— to be produced by a sequence of production states, $S_{1,m}, \ldots, S_{T,m}$, during a given time frame (namely the $m^{th}$ of $M$ time frames). However, in addition we now assume that the way in which the sequence of production states is generated depends on which of $K$ possible internal states is active during the current time frame. The length of the sequence of production states produced by the $k^{th}$ internal state can be dictated by the data collection process or imposed by the analysis. The corresponding $K$-state internal state process, $\{H_m\}_{m=1}^M$, is such that $H_m$ serves as a proxy for a behavior occurring at a cruder time scale, namely throughout the $m^{th}$ time frame. In this manner, we account for differences observed across the time series $\mathbf{y}_{m}$, $m=1,\ldots,M$, by connecting each with one of $K$ crude-scale behavioral processes while still modeling the transitions among production states at the time scale at which the data were collected. Supposing that there are multiple time series per individual, we can model the manner in which an animal switches among the $K$ internal states (behavioral processes). 
We assume a Markov process at the individual time series level, i.e.\ $\Pr (H_m | H_{m-1},\ldots,H_1) = \Pr (H_m| H_{m-1})$, such that the $m^{th}$ internal state is conditionally independent of all other internal states given the internal state at the $(m-1)^{th}$ time point. The $K \times K$ t.p.m.\ for the internal states $\{H_m\}_{m=1}^{M}$ examines persistence in the internal states, as well as the manner in which an animal will switch among them. Figure 1 displays the dependence structure of hierarchically structured HMMs with two Markov chains, one at the level of the production states, $S_{t,m}$, and the other at the level of the internal states, $H_m$. 

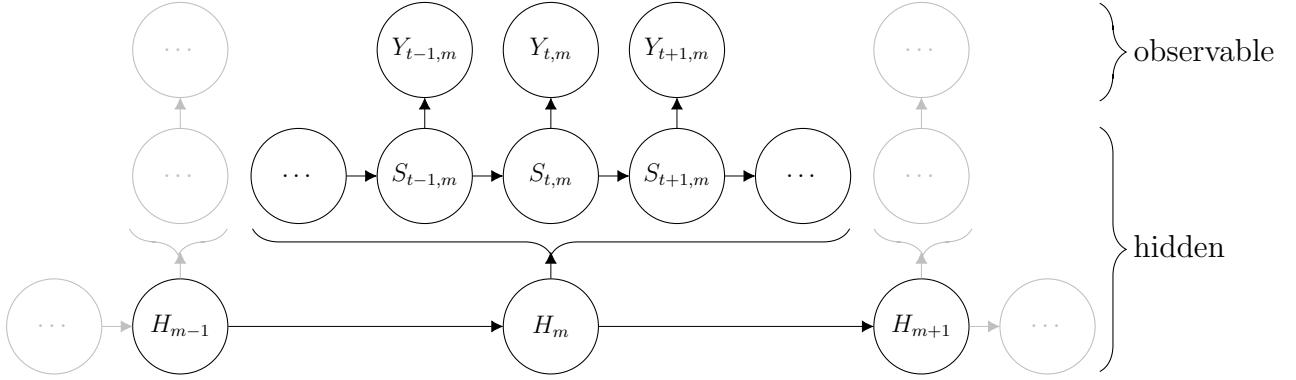
\begin{figure}
\centering
\begin{tikzpicture}[node distance = 1.5cm]
\tikzset{state/.style = {circle, draw, minimum size = 45pt, scale = 0.80}}
\node [state, color=lightgray] (0) {$\cdots$};
\node [state] (1) [right = 4mm of 0] {$H_{m-1}$};
\node [state] (2) [right = 36.5mm of 1] {$H_m$};
\node [state] (3) [right = 36.5mm of 2] {$H_{m+1}$};
\node [state, color=lightgray] (4) [right = 4mm of 3] {$\cdots$};
\node [state] (6) [above = 7.25mm of 2] {$S_{t,m}$};
\node [state] (5) [left = 4mm of 6] {$S_{t-1,m}$};
\node [state] (7) [right = 4mm of 6] {$S_{t+1,m}$};
\node [state] (8) [above = 4mm of 5] {$Y_{t-1,m}$};
\node [state] (9) [above = 4mm of 6] {$Y_{t,m}$};
\node [state] (10) [above = 4mm of 7] {$Y_{t+1,m}$};
\node [] (14) [above = 3.5mm of 2] {};
\node [state] (15) [left = 4mm of 5] {$\cdots$};
\node [state] (16) [right = 4mm of 7] {$\cdots$};
\node [state, color=lightgray] (17) [above = 7.25mm of 1] {$\cdots$};
\node [state, color=lightgray] (18) [above = 7.25mm of 3] {$\cdots$};
\node [state, color=lightgray] (19) [above = 4mm of 17] {$\cdots$};
\node [state, color=lightgray] (20) [above = 4mm of 18] {$\cdots$};
\node [text=black,] (11) [right = 20.4mm of 20] {observable};
\node [text=black,] (12) [below = 20.2mm of 11] {\hspace*{-8mm} hidden};
\node [] (21) [above = 3.5mm of 1] {};
\node [] (22) [above = 3.5mm of 3] {};
\draw[->, color=lightgray, line width=0.2pt] (0) to (1);
\draw[->, black, line width=0.2pt] (15) to (5);
\draw[->, black, line width=0.2pt] (7) to (16);
\draw[->, black, line width=0.2pt] (1) to (2);
\draw[->, black, line width=0.2pt] (2) to (3);
\draw[->, color=lightgray, line width=0.2pt] (3) to (4);
\draw[->, black, line width=0.2pt] (5) to (6);
\draw[->, black, line width=0.2pt] (6) to (7);
\draw[->, black, line width=0.2pt] (2) to (14);
\draw[->, black, line width=0.2pt] (5) to (8);
\draw[->, black, line width=0.2pt] (6) to (9);
\draw[->, black, line width=0.2pt] (7) to (10);
\draw[->, color=lightgray, line width=0.2pt] (1) to (21);
\draw[->, color=lightgray, line width=0.2pt] (3) to (22);
\draw[->, color=lightgray, line width=0.2pt] (17) to (19);
\draw[->, color=lightgray, line width=0.2pt] (18) to (20);
\draw[decorate, decoration={brace, amplitude=10pt}] (10.575,1.3) -- (2.6375,1.3) node (6)[midway, below,]{};
\draw[decorate, decoration={brace, amplitude=10pt}, color=lightgray] (2.3,1.3) -- (1.0,1.3) node (6)[midway, below,]{};
\draw[decorate, decoration={brace, amplitude=10pt}, color=lightgray] (12.2,1.3) -- (10.9,1.3) node (6)[midway, below,]{};
\draw[decorate,decoration={brace,amplitude=10pt}] (13.9,2.65) -- (13.9,-0.6) node (6)[midway,below,]{};
\draw[decorate,decoration={brace,amplitude=10pt}] (13.9,4.3) -- (13.9,3.0) node (6)[midway,below,]{};
\end{tikzpicture}
\caption{Dependence structure in hierarchically structured HMMs.}
\end{figure}

Let $\boldsymbol{\delta}^{(I)}$ denote a $K-$vector of initial probabilities for the internal states, let $\boldsymbol{\Gamma}^{(I)}$ denote the $K \times K$ t.p.m.\ for the internal state process, and define 
$$\mathbf{P}^{(I)}(\mathbf{y}_m) = \text{diag} \bigl( L_p (\mathbf{y}_{m}|H_m=1), \ldots, L_p (\mathbf{y}_{m}|H_m=K) \bigr).$$

The likelihoods $L_p (\mathbf{y}_{m}|H_m=k)$, $k=1,\ldots, K$, have the form as given in (\ref{basiclikelihood}), and only vary across $k$ in terms of the implied production-level t.p.m.\ (and potentially also the production state-dependent distributions). Then the likelihood for the hierarchically structured HMM is, 
\begin{linenomath}
\begin{equation*}
L_{h}= \boldsymbol{\delta}^{(I)} \mathbf{P}^{(I)}(\mathbf{y}_1) \prod_{m=2}^M \boldsymbol{\Gamma}^{(I)} \mathbf{P}^{(I)}(\mathbf{y}_m) \mathbf{1}
\end{equation*} 
\end{linenomath}
As the estimated production states are generally proxies for behaviors, allowing for only the t.p.m.\ to vary across the $K$ HMMs leads to an interpretation of the $K$ internal states (loosely connected to $K$ behavioral processes) as distinct manners in which an animal will persist and switch among the production states. As long as the individual time series' likelihoods, $L_p$, can be evaluated in an efficient manner, we can evaluate the likelihood of the hierarchically structured HMM via the forward algorithm, since the general structure does not differ from that of the basic HMM, and thus maximize it directly \citep{zuc16}. The Viterbi algorithm can be used for global state decoding, i.e.\ finding the sequence of the most likely internal and production states, respectively, given the observations.

\section{APPLICATIONS} \label{sec:apps}

\subsection{HARBOR PORPOISES}

\textit{3.1.1 The data}. To illustrate the application of hierarchically structured HMMs, we model vertical movements of a harbor porpoise (\textit{Phocoena phocoena}) throughout its natural habitat in the northeastern part of the North Sea. From a time-depth recorder (LAT1800ST, Lotek, Ontario, Canada), we obtained observations of the dive depth every second. Assuming a ``dive'' to be any vertical movement deeper than two meters below the surface, we used the \texttt{R} package \texttt{diveMove} \citep{luq07} to process
the raw data into measures of the dive duration, the maximum depth and the dive wiggliness (as represented by the absolute vertical distance covered at the bottom of each dive) to characterize the porpoise's vertical movements at a dive-by-dive resolution. Previous applications of HMMs, though not hierarchically structured HMMs, with dive-by-dive data of marine mammals have been presented in \citet{har10} and \citet{der16}. Overall, we consider 275 hours of observations, comprising 7,585 dives in total (hence, about 28 dives per hour).

\noindent \textit{3.1.2 Model formulation and model fitting}.
Behavioral modes of marine mammals, e.g., resting, foraging and traveling, do not necessarily manifest themselves at a dive-by-dive resolution.
For example, foraging behavior typically coincides with a large proportion of extensive, wiggly dive sequences. However, foraging sequences may be interspersed by short periods of resting behavior (shallow and smooth dives) even though the dominant behavioral mode may still be foraging. 
Such patterns are especially likely to occur in harbor porpoise dive data, a species that needs to feed almost continuously to meet energy requirements \citep{wis16}. In these cases, hierarchically structured HMMs have strong potential to infer the movement strategies adopted over time, by modeling the transitions between distinct dive patterns (as represented by multiple HMMs) rather than modeling dive-by-dive observations using a single HMM. Thus, to draw a more detailed picture of the behavioral dynamics at multiple time scales, we use hierarchical HMMs, where a crude-scale $K$-state Markov chain selects which of $K$ fine-scale HMMs describes the dive pattern observed at any point in time. Intuitively, the crude-scale process describes the general behavioral mode (e.g.\ resting or traveling) — which may persist for a large number of consecutive dives — while the fine-scale process captures more nuanced state transitions at the dive-by-dive level, given the general behavioral mode.

In terms of the crude time scale, we segmented the time series into hourly intervals and allowed each segment to be connected to one of $K=2$ HMMs with $N=3$ (dive-by-dive-level) states each. This somewhat arbitrary time scale was chosen based on exploratory analysis of the data set, which suggested that a certain dive pattern is typically adopted for several hours before switching to another one (c.f.\ Figure \ref{fig5}). As comprehensively discussed in \citet{poh17}, model selection criteria such as Akaike's Information Criterion (AIC) or the Bayesian Information Criterion (BIC) typically tend to favor models with larger number of states than are biologically sensible. This is indeed a well-known and notorious problem in applications of HMMs to ecological data (see also \citealp{lan15}, \citealp{der16}, \citealp{li17}). Thus, following \citet{poh17}, instead of relying on formal model selection procedures, the number of states was chosen pragmatically, with particular emphasis on model parsimony and biological intuition.

The state-dependent distributions were kept the same across the two dive-level HMMs, which were instead allowed to differ only by the t.p.m.s. This assumption implies that any of the three types of dives --- as generated by the three different production states --- could in principle occur in both crude-level behavioral modes, but will not occur equally often, on average, due to the different Markov chains active at the dive-by-dive level.
The initial state distributions, both for the internal and for the production state process, were assumed to be the stationary distributions of the respective Markov chains.
We assumed gamma distributions for each of the three dive variables (dive duration, maximum depth and dive wiggliness), with an additional point mass on zero in case of dive wiggliness to account for the zeros observed. We assumed contemporaneous conditional independence, 
i.e.\ for any given dive, the three variables observed are conditionally independent given the production state active at the time of the dive \citep{zuc16}. These assumptions could in fact be relaxed if deemed necessary. However, for this case study we decided that in order to illustrate the key concepts, it would be best to focus on a relatively simple yet  biologically informative model structure.

We computed the likelihood using the forward algorithm \citep{zuc16} and used the R function \texttt{nlm} \citep{r16} to obtain maximum likelihood estimates via direct numerical likelihood maximization.

\begin{figure}[htb!]
\centering
\includegraphics[width=\textwidth]{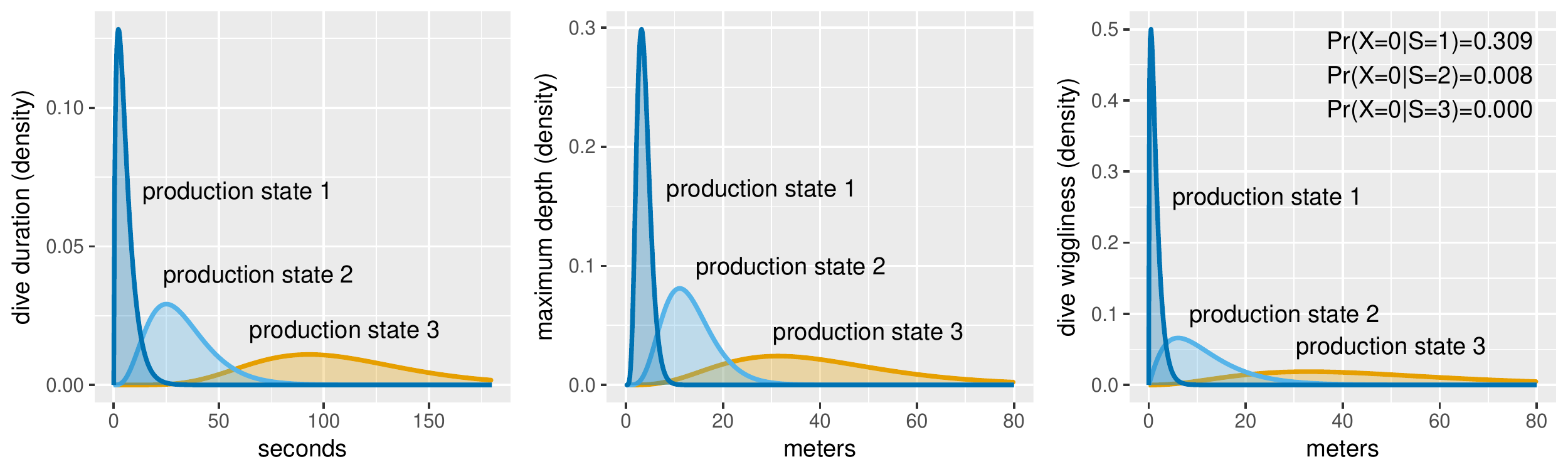}
\caption{Fitted state-dependent distributions for the dive duration, the maximum depth and the dive wiggliness, the latter together with the estimated point mass on zero.}
\label{fig4}
\end{figure}

\begin{figure}[t!]
\centering
\includegraphics[width=\textwidth]{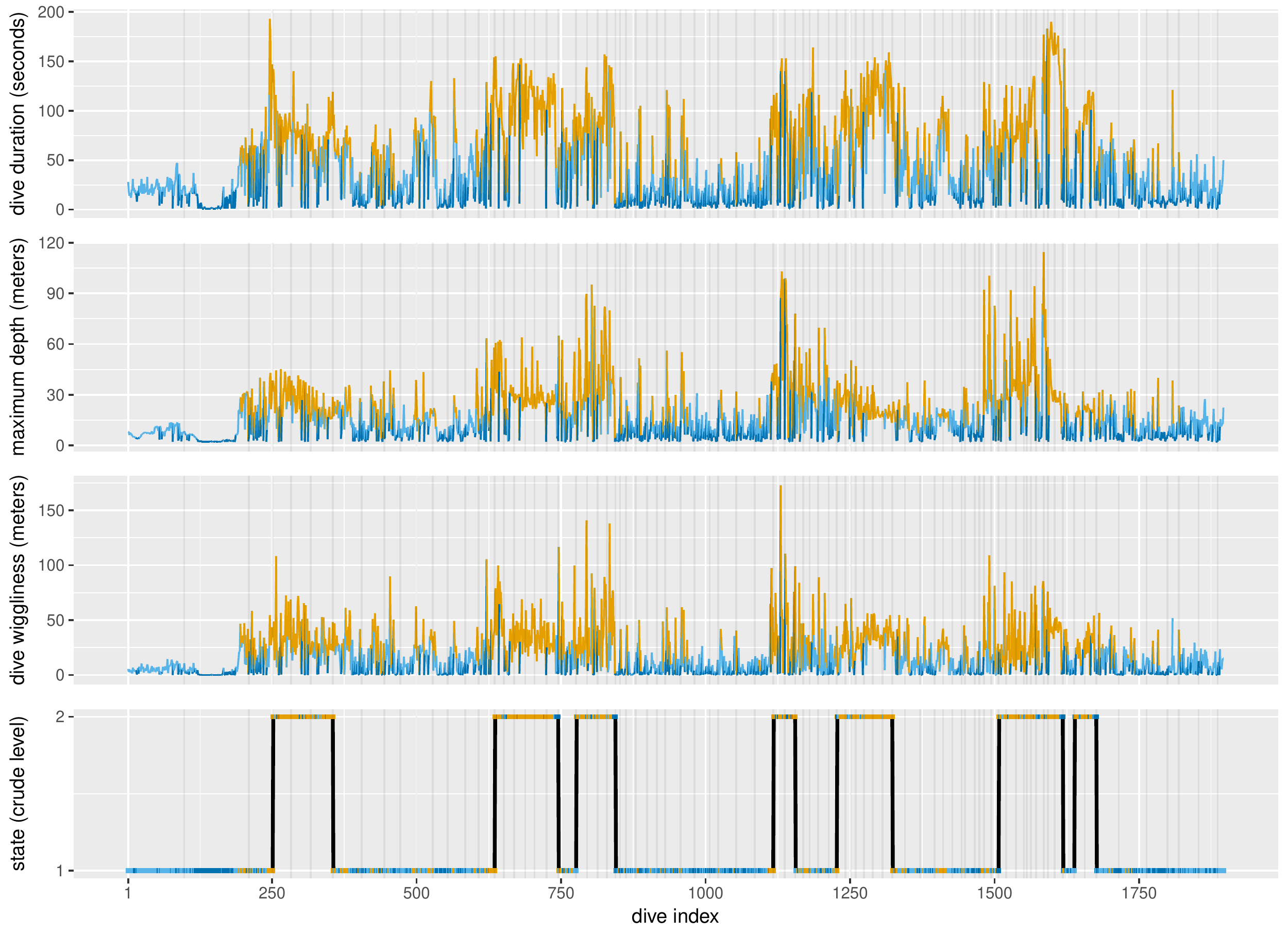}
\caption{Exemplary sequence of the first 25\% of the observations of the dive duration, the maximum depth and the dive wiggliness. Hourly segments are indicated by vertical grey lines. The states were decoded using the Viterbi algorithm.}
\label{fig5}
\end{figure}

\noindent \textit{3.1.3 Fitted state-dependent distributions}. The fitted (dive-level) state-dependent distributions displayed in Figure \ref{fig4} suggest three distinct dive types:
State 1 captures the shortest (lasting less than 25 seconds), shallowest (less than 10 meters deep) and smoothest (less than 8 meters absolute vertical distance covered) dives with small variance.
State 2 captures moderately long (10-60 seconds), moderately deep (5-25 meters) and moderately wiggly (5-30 meters) dives with moderate variance.
State 3 captures the longest (40-180 seconds), deepest (10-80 meters) and wiggliest (10-80 meters) dives with high variance.

In the next section, we discuss the ($K=2$) distinct dive-level switching patterns among the ($N=3$) states discussed here, as well as the crude-level process that selects which of the dive-level switching patterns is active in any given hour.

\noindent \textit{3.1.4 Estimated transition probability matrices}. The t.p.m.\ of the crude-level Markov chain, which selects among the dive-level HMMs, and the t.p.m.s of those two dive-level HMMs, which describe the switching between different types of dives, were estimated as follows:
\begin{itemize}
\item crude level: \[
\hat{\boldsymbol{\Gamma}}^{(I)} = \begin{pmatrix}
0.789 & 0.211 \\
0.219 & 0.781
\end{pmatrix}
\]
\item dive level: \[
\hat{\boldsymbol{\Gamma}}_1^{} = \begin{pmatrix}
0.406 & 0.443 & 0.150\\
0.240 & 0.600 & 0.159\\
0.196 & 0.366 & 0.437
\end{pmatrix} \text{ and } ~
\hat{\boldsymbol{\Gamma}}_2^{} = \begin{pmatrix}
0.277 & 0.153 & 0.570\\
0.124 & 0.248 & 0.628\\
0.057 & 0.087 & 0.856
\end{pmatrix}
\]
\end{itemize}
The corresponding stationary distributions are $(0.509, 0.491)$, $(0.277, 0.506, 0.217)$ and $(0.083,$ $0.110, 0.807)$, respectively. The former of these three stationary distributions implies that, according to the fitted model, in the long run, approximately half of the observations were generated by each of the two HMMs. Furthermore, according to the estimated t.p.m.\ $\hat{\boldsymbol{\Gamma}}^{(I)}$, there is fairly strong persistence in the crude-level states, indicating that the porpoise typically remains in any given internal state for several hours before switching to the other internal state. This is also confirmed by Figure \ref{fig5}, which displays the first 25\% of the decoded observations. In particular, Figure \ref{fig5} shows that there are bouts of several hours where production states 1 and 2 are dominant (yet still interspersed with occasional dives generated by production state 3), but also such where production state 3 is dominant. Bouts of the former type are assigned to internal state 1, while the latter are assigned to internal state 2. This again highlights the need to apply hierarchically structured HMMs, here effectively as a means to account for temporal heterogeneity in the state-switching pattern exhibited by the porpoise.

At the dive level, when the first HMM is active, then in the long run about 28\%, 51\% and 22\% of the observations are generated in state 1, 2 and 3, respectively, whereas when the second HMM is active, in the long run about 8\%, 11\% and 81\% of the dives are generated in the respective states. Furthermore, if the first HMM is active, then switching takes place primarily between states 1 and 2, and additionally from state 3 to state 2.
If the second HMM is active, then state 3 is dominant, with fairly strong persistence and lots of switches from states 1 and 2 to state 3.

\noindent \textit{3.1.5 Concluding remarks}. The second HMM is indicative of foraging behavior, particularly due to the extensive wiggliness during dives, which often indicates prey-chasing. The interpretation of the first HMM, which involves a large proportion of relatively short, shallow and smooth dives, could indicate a resting and/or a traveling behavior. Indeed, traveling from one area to another while remaining close to the water surface is likely the most efficient strategy. However, a more detailed interpretation of the first HMM would require inclusion of other variables such as the step length, which may prove useful to distinguish between resting and traveling.

\subsection{GARTER SNAKES}

\noindent \textit{3.2.1 The data}. We model the movements of 19 juvenile garter snakes ({\it Thamnophis elegans}) in repeated trials that are a subset from a larger experiment quantifying behaviors in the offspring of wild-caught females across experimental treatments, manuscript in preparation. Using EthoVision XT 8.5  (Noldus Information Technology, Wageningen, The Netherlands), we extracted movement data for each of the  snakes across three trials. In brief, snakes were placed in a novel test arena (circular enclosure with diameter of 24.5 cm) for 120 s. Trials were videorecorded and divided into two segments, each lasting 50 s, to account for any possible disturbances of snakes by observer movement at the beginning and end of each trial. This resulted in a total of six recorded segments for each snake. The individuals included here represent the control group, which was not exposed to any additional in the test arena during the first two trials and was exposed to a novel object after the first minute of the third trial (that is, between tracks 5 and 6).

\noindent \textit{3.2.2 Model formulation and model fitting}. The snakes displayed a variety of general movement strategies, from the extreme of remaining motionless to moving rapidly around the test arena for the duration of the trial. We calculated the distance moved within 1/2 s and subsequently applied a square root transformation to deal with extreme values present. We assumed that each observed distance conditional on one of three production states was generated by a state-dependent gamma density. Further, to investigate habituation and behavioral plasticity over the course of the six time series per snake, we assumed that each time series was generated by one of three internal state-dependent HMMs. The complete hierarchically structured HMM fitted to the observed distances was composed of three production states, kept the same across the internal states, and three internal states. In this manner, we investigated whether there was persistence at the internal state level, i.e.\ if the garter snakes tended to repeat the same general movement patterns across time series or switch strategies.

\begin{figure}[h!] 
\centering
\includegraphics[scale=.65]{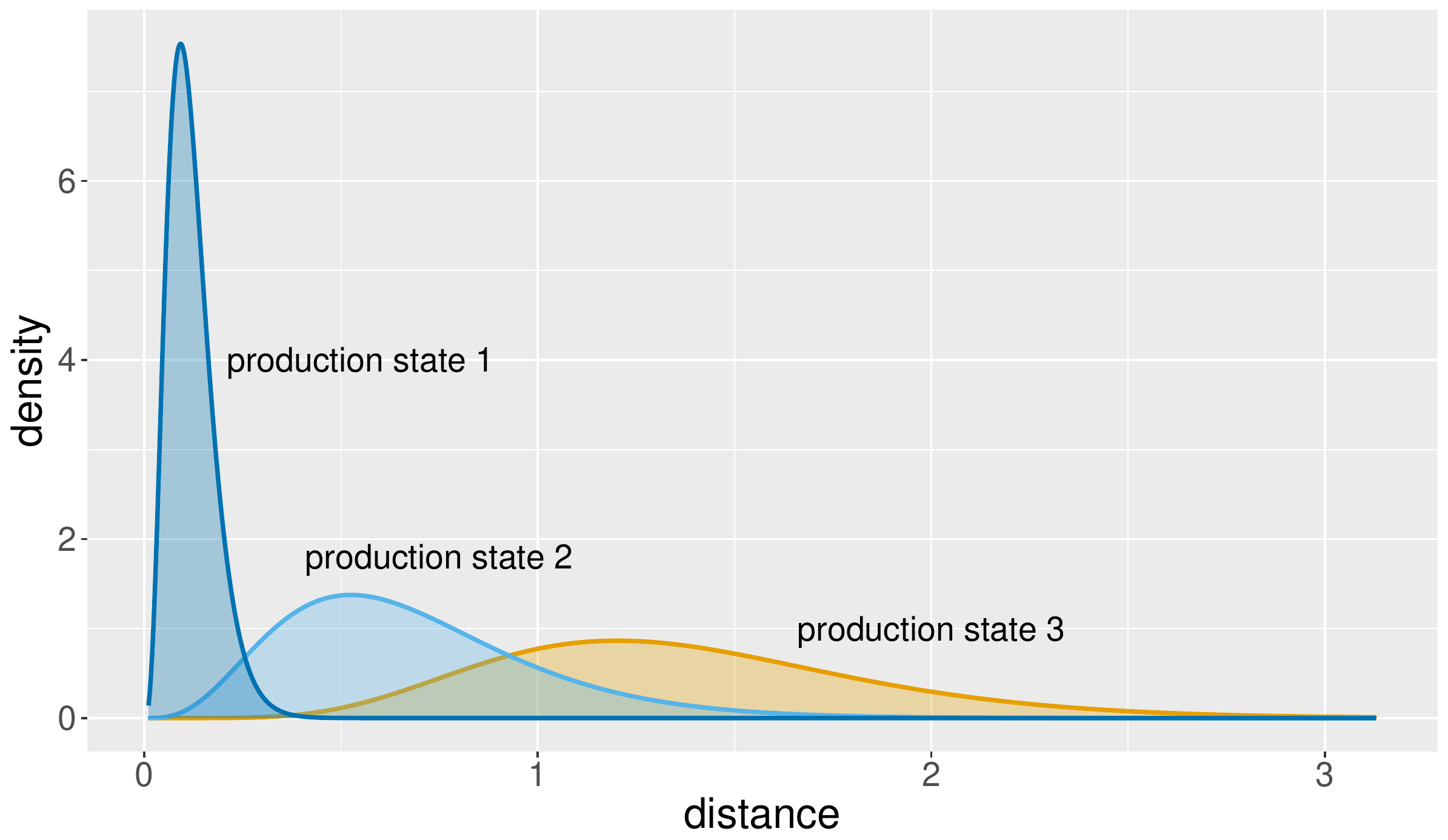}
\caption{Fitted state-dependent distributions for distance traveled.}
\label{fig:gssdd}
\end{figure}

\noindent \textit{3.2.3 Fitted state-dependent distributions}. The fitted state-dependent gamma distributions for the three production states, shown in Figure \ref{fig:gssdd}, correspond to three general types of movement strategies: motionless (or nearly so), slow exploratory, and rapid escape, which the video recordings demonstrate. The estimated average distance traveled in production states 1--3 are:  0.0148, 0.459 and 1.891 cm per 1/2 s, respectively. The largest amount of variability in observed step lengths corresponds to production state 3, with a standard deviation of 0.487 cm$^{1/2}$. 

\noindent \textit{3.2.4 Estimated transition probability matrices and initial state distributions}.

\begin{itemize}
\item crude level: \[
\hat{\boldsymbol{\Gamma}}^{(I)} = \begin{pmatrix}
0.166 & 0.578 & 0.256 \\
0.680 & 0.226 & 0.095 \\
0.157 & 0.208 & 0.635 \\
\end{pmatrix} \text{   } ~
\hat{\boldsymbol{\delta}}^{(I)} = \begin{pmatrix}
0.903 \\
0.072 \\
0.025
\end{pmatrix}
\]
\item movement level: 
\[
\hat{\boldsymbol{\Gamma}}_1^{} = \begin{pmatrix}
0.947 & 0.047 &  0.006\\
0.018 & 0.919 & 0.063\\
\sim 0 & 0.244 & 0.756
\end{pmatrix} \text{   } ~
\hat{\boldsymbol{\delta}}_1^{} = \begin{pmatrix}
0.413\\ 
0.103 \\ 
0.484 \\
\end{pmatrix}
\]
\[
\hat{\boldsymbol{\Gamma}}_2^{} = \begin{pmatrix}
0.806 & 0.144 & 0.050\\
0.019 & 0.657 & 0.324\\
\sim 0 & 0.185 & 0.815
\end{pmatrix} \text{   } ~
\hat{\boldsymbol{\delta}}_2^{} = \begin{pmatrix}
0.087 \\ 
0.024 \\ 
0.889
\end{pmatrix}
\]
\[
\hat{\boldsymbol{\Gamma}}_3^{} =\begin{pmatrix}
0.994 & 0.006 & \sim 0\\
0.003 & 0.997 & \sim 0\\
\sim 0 & 0.018 & 0.982
\end{pmatrix} \text{   } ~
\hat{\boldsymbol{\delta}}_3^{} = \begin{pmatrix}
0.315 \\ 
0.442 \\ 
0.243\\ 
\end{pmatrix}
\]
\end{itemize}

The three estimated t.p.m.s at the movement level, corresponding to the three internal states, can generally be interpreted as representing three different levels of behavioral flexibility. When the first internal state is active, characterized by $\hat{\boldsymbol{\Gamma}}_1^{}$, individuals are showing more persistence in the motionless and slow exploratory behavioral states overall, and are likely to transition from the rapid escape state to the exploratory state. When the second internal state is active, $\hat{\boldsymbol{\Gamma}}_2^{}$ demonstrates that individuals are switching regularly between behavioral states. When the third internal state is active, $\hat{\boldsymbol{\Gamma}}_3^{}$ demonstrates that individuals seldom transition between states. Thus, the three internal states reflect a continuum of behavioral flexibility within a short, but ecologically relevant time scale in the context of a potential predation event ($<$ 1 min). 
\begin{figure}[h!]
\centering
\includegraphics[scale=.7]{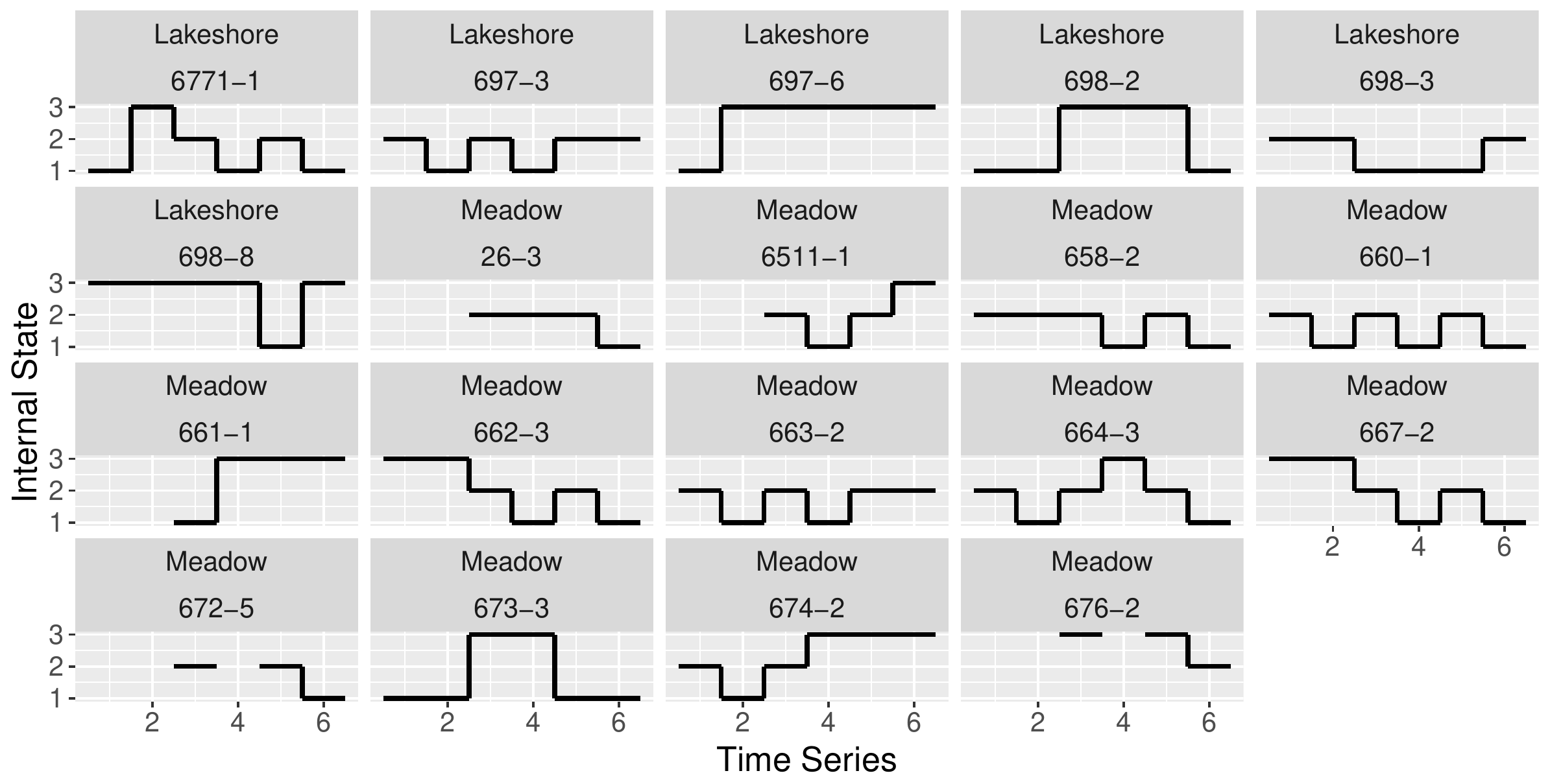}
\caption{Internal state decodings by garter snake. Each garter snake is from one of two ecotypes: Lakeshore or Meadow.} 
\end{figure}

At the crude level, the t.p.m.\ $(\hat{\boldsymbol{\Gamma}}^{(I)}$) indicates that individuals are readily switching among the three movement level HMMs across trials and in fact are more likely to switch between the movement HMMs  representing the greatest behavioral flexibility $(\hat{\boldsymbol{\Gamma}}_1^{}$ and $\hat{\boldsymbol{\Gamma}}_2^{})$. At the crude time scale, we observe persistence in the the movement-level HMM describing snakes that are behaviorally inflexible in their movements $(\hat{\boldsymbol{\Gamma}}_3^{}$). Overall, these results indicate that, at the broader time scale, many individuals are readily altering their level of behavioral flexibility while some individuals remain persistent in their behaviors both within and across trials (i.e., at both the movement and crude levels).

\noindent \textit{3.2.5 Concluding remarks}.
HMMs are not yet commonly applied to animal movement data from experimental designs, even though they typically produce multiple time series per individual. Introducing multiple Markov chains in the HMM formulation lends itself to characterizing the consistency of individual behaviors and variation among individuals at different time scales. We show that individuals employing multiple movement strategies in a narrow time frame are more likely to switch between strategies at a crude time scale, while individuals consistent in their behaviors at the movement time scale are also consistent at the crude time scale. Furthermore, these patterns are independent of the behavioral strategies exhibited: individuals may consistently remain in any one of the three behavioral states.

\section{DISCUSSION}

HMMs have proven to be useful statistical tools for modeling animal movement data, providing a framework to infer drivers of variation in movement patterns, and thus behavior. The basic HMM, however, has so far been used to infer aspects of animal behavior only when a single data point can be thought to stem from one of $N$ possible (production) states, which are loosely connected to behavioral modes that manifest themselves at the temporal resolution at which observations are made. Yet, thanks to advances in tag technology and battery life, data can be collected at finer temporal resolutions and over longer periods of time. Inferences at time scales cruder than those at which data are collected, and which correspond to larger-scale behavioral processes, are not yet answered via HMMs. We provide a corresponding extension to incorporate multiple Markov chains in an HMM, allowing for multi-scale behavioral inferences. The extension is straightforward in the sense that likelihood inference via application of the forward algorithm is essentially analogous as in case of basic HMMs. The hierarchically structured HMMs can also be used to avoid coarsening data, such as acceleration data that can be collected many times per second \citep{leo16}. As this is, as of yet, an area of movement ecology that has received little attention, our proposed framework is one of the first that models animal behavior simultaneously at multiple time scales. 

In this manuscript, we did not discuss how to implement model selection and model checking for hierarchically structured HMMs. In principle, since we are fitting the models using maximum likelihood, model selection could be conducted using standard information criteria. However, while conceptually this is completely straightforward, in practice this procedure is notoriously error-prone already for basic HMMs, due to the strong tendency of information criteria to favor models with many more states than are biologically reasonable \citep{lan15,poh17,li17}. Given the additional state process, this issue will be exacerbated within hierarchically structured HMMs as presented in this work, since the number of states both for the production process and for the internal state process needs to be chosen. We cannot currently offer a satisfactory solution to this problem, except by saying that biological {\it a priori} expert knowledge ought to be taken into account. For general advice regarding the issue of model selection in HMMs, see \citep{poh17}. For model checking, possible avenues are (i) simulation-based model assessment and (ii) analyses of pseudo-residuals. Regarding (i), the fundamental concept is the idea that the fitted model should generate data similar to the observed data in all important aspects. Quantification of aspects of the data patterns should reflect key behaviors believed to be important to the problem. Pseudo-residuals, as discussed for example in \citet{pat09}, \citet{lan12} and in \citet{zuc16}, can be calculated also for hierarchically structured HMMs, most easily by conditioning on Viterbi-decoded internal states, hence calculating the pseudo-residuals at the production level, given the (fixed) most likely internal state sequence. Both model selection and model checking needs to be explored further before these models may become a tool that is routinely applied in the analysis of animal behavior data.

Using {\it ad hoc} choices of the exact model formulations (yet such that are grounded in biological theory), in Section \ref{sec:apps} we demonstrated how the hierarchically structured HMMs, applied to movement data collected on harbor porpoises and garter snakes, respectively, provided new insights into the behavior of these species. However, a hierarchically structured HMM not only allows for new inferences to be made from movement studies --- it can also be applied to the study of behavior in general. Being able to characterize persistence of movement patterns at multiple time scales allows us to learn about personality, individual specialization, and cognition, among other things. Several studies across a wide range of taxa have shown that individual animals behave differently from other individuals and that these differences are maintained through time \citep{rea07, sih04, din05, bir08}. These observations have given rise to the burgeoning field of animal personality which explores the ecology and evolutionary significance of such behavioral differences among individuals. Such studies have included a variety of behavioral measures but have only recently incorporated models of movement as a behavioral trait \citep[e.g.][]{sch12, mck15, spi17}. Importantly, the animal personality framework has recently incorporated an understanding of how individuals differ in their behavioral plasticity \citep[reviewed in][]{mat15, sta16}, which requires more specific theoretical models as well as more sophisticated statistical approaches  \citep{din13,kle13,jap14}. Thus, the field is attempting to address two fundamental questions: (1) how do behaviors differ among individuals and (2) how do individual behaviors change over time or context? Addressing these questions therefore requires analysis at two levels: (1) to identify and categorize behavioral states (production states) and (2) to identify patterns of changes in behavioral states (internal states). In the HMM framework, the internal states may reflect general movement patterns associated with endogenous behavioral plasticity \citep[\textit{sensu}][]{sta16} or personality which allows for further examination of persistence or switching among them at the cruder time-scale. 

The addition of multiple Markov chains in the HMM framework to conduct multi-scale behavioral inferences necessitates the selection of the temporal resolution at two time scales: the observation level and the level of the individual time series. The selected temporal resolution at the level of the internal states will need to be tied to the specific biological question of interest. There may be a natural manner in which the data are segmented that produces time series of unequal length. However, this need not be an issue as long as each time series is reflective of some general behavioral process irrespective of the length of the time series. Formulating the hierarchically structured HMM, in terms of selecting the number of production states and internal states, will need to be done in a pragmatic fashion in order to balance model complexity with biological intuition. Due to the HMM's inherent flexibility, the internal states may be formulated in a few manners, e.g.\ a single HMM (such as has been described in Section 2), assuming a distribution of HMMs, or allowing for longer state dwell times via the hidden semi-Markov model, in order to account for unexplained variability in the state processes. In particular, as the number of production states, $N$, increases, so will the number of ways in which two HMM's t.p.m.s $\boldsymbol{\Gamma}_i$ and $\boldsymbol{\Gamma}_j$ can differ. To account for all of these possibilities may require a large number of internal states, if each internal state is assumed to only correspond to one t.p.m.\ for the HMM.

Adding hierarchical structures to the HMM opens new possibilities for modeling multi-scale behaviors and provides an avenue to study animal personality and general behavior from movement studies. In this manner, environmental covariates can also be included to understand their effects on state occupancy and dynamics of variation in behavioral modes at broader time-scales than that at which the data are collected. Further, this framework may be adapted for simultaneous modeling of multiple animal behavior data streams collected at distinct temporal resolutions. The internal states can be adapted to generate a sequence of fine-scale observations as well as one observation from a distinct data stream.

\section{ACKNOWLEDGEMENTS}

The harbor porpoise movement data were collected as part of the DEPONS project \linebreak (www.depons.au.dk) funded by the offshore wind developers Vattenfall, Forewind, SMart Wind, ENECO Luchterduinen, East Anglia Offshore Wind and DONG Energy.
Funding for snake project provided by Iowa Science Foundation (15-11) and the Gaige Award of the American Society of Ichthyologists and Herpetologists. EJG partially supported by a fellowship from the ISU Office of Biotechnology.


\renewcommand\refname{REFERENCES}
\makeatletter
\renewcommand\@biblabel[1]{}

\end{spacing}
\end{document}